# Consistent and fault-tolerant SDN with unmodified switches

André Mantas (*student)*

amantas@lasige.di.fc.ul.pt

Fernando M. V. Ramos

fvramos@ciencias.ulisboa.pt

LaSIGE, Faculdade de Ciências, Universidade de Lisboa, Portugal

In Software-Defined Networking (SDN), applications use the logically centralized network view provided by the controllers to program switches in the network.

**Motivation.** In a reliable SDN environment, different controllers coordinate different switches and backup controllers can be set in place to tolerate faults. This approach increases the challenge to maintain a consistent network view. Indeed, if this global view is not consistent with the actual network state, applications will operate on a stale state and potentially lead to incorrect behavior. These anomalies can degrade network performance and introduce problems such as network loops or security breaches [1].

To avoid this problem, controllers need to maintain a consistent network view even in the presence of faults. In order to build a consistent view, network events (packets received by controllers from the network) need to be processed in a consistent way, guaranteeing three properties: (i) events are processed in the same (total) order in all controllers, (ii) no events are lost (processed *at least once*) and (iii) no events are processed repeatedly (*at most once*). These properties ensure that all controllers will reach the same internal state and thus build a consistent network view.

However, maintaining consistent controller state is not enough. In SDN, it is necessary to include switch state into the system model and handle it consistently [2]. Mechanisms need to be put in place to ensure that controllers actually send commands to the switches *exactly once*. This form of consistency (in both controllers and switches) is needed to offer full transparency to network applications and make controllers more robust.

Ravana [2] was the first controller to provide this level of consistency in the control plane. For this purpose, Ravana modifies the OpenFlow (OF) protocol and makes changes to current switches. Namely, it requires explicit acknowledgements (ACKs) to be added to the protocol, and it leverages on buffers implemented on switches to retransmit events and filter possible repeated commands.

**Problem.** Ravana is an elegant solution, but requiring changes to OF and to switches hinders its adoption. First, it is necessary an agreement for a protocol to change. And even if that first step is granted, it can take significant time for changes to be introduced. Second, despite the exciting new trend in data plane programmability, it is implausible that commodity switches will enable Ravana requirements soon.

**Contribution.** Our proposal is a fault-tolerant controller that gives the same guarantees as Ravana – exactly-once events and exactly-once commands – *without* the need to modify the protocol or switches.

**Exactly-once events:** Similar to Ravana, controllers still coordinate their state using a state machine replication protocol. However, switch events are sent to *all* controllers (primary and backups) using the master/slave model, with slaves registering to receive all asynchronous messages. In Ravana events are only sent to the primary (which is enough due to the use of ACKs and buffering of events in the switch).

**Exactly-once commands:** The reception of commands is acknowledged using OF Bundles (introduced in version 1.4). A bundle is a sequence of OF modification requests from the controller that is applied as a single OF operation. To ensure that a new master does not send repeated commands, *all* controllers are notified when a bundle is committed on the switch. This is a challenge as, according to the standard, the bundle commit reply is sent in the same connection of the request (i.e., to the master). To tackle it, we consider two solutions: (a) adding PacketOut messages to the bundle that will be forwarded to the slave controllers when the bundle is committed (the OF specification allows this, but some switches do not yet support it) and (b) configure the switches' internal forwarding state (used to forward control packets) to send these messages to all controllers via proprietary commands (this is supported by Pica8 switches, for instance). Again, these techniques do not require changes to switches or OF.

**Future.** We are implementing a prototype using Floodlight, which will be released as open source. The system will be evaluated and compared to Floodlight and Ravana in the absence and presence of faults to evaluate the impact of the increased number of control messages that results from our solution.

This project has received funding from the European Union's Horizon 2020 research and innovation programme under grant agreement No H2020-643964 (SUPERCLOUD), and by national funds through Fundação para a Ciência e a Tecnologia (FCT) with reference UID/CEC/00408/2013 (LaSIGE).